\documentstyle[aps,preprint,epsf]{revtex}
\begin{document}
\draft
\title{Conductance Peak Distributions  in Quantum Dots at Finite
 Temperature: Signatures of the Charging Energy}

\author{ Y. Alhassid,   M. G\"{o}k\c{c}eda\u{g} and A.D.  Stone}

\address{ Center for Theoretical Physics, Sloane Physics Laboratory,
	 Yale University, New Haven, Connecticut 06520, USA }

\date {submitted July 17, 1997}
\maketitle
\begin{abstract}
 We derive the finite temperature conductance peak distributions and 
peak-to-peak correlations for quantum dots in the Coulomb blockade regime
 assuming the validity of random matrix theory.
 The distributions are universal, depending only on the symmetry
 class and the temperature measured in units of the mean level spacing, 
$\Delta$.
When the temperature is comparable to $\Delta$ several resonances
 contribute to the same conductance peak and we find
 significant deviations from the previously known $T \ll \Delta$ distributions.
In contrast to the $T \ll \Delta$ case, these distributions 
show a strong signature of the charging energy and charge quantization
on the dot. 

\end{abstract}

\narrowtext

  Quantum dots are two-dimensional microstructures of micron scale
 or smaller in which a small number
of electrons are confined by electrostatic potentials. They can
 be fabricated  with relatively
little intrinsic disorder, in which case the motion of the electrons is
ballistic.  The transport properties of such dots are
measured by coupling them to leads through point contacts. As the
 contacts are pinched off, the dot becomes more closed  (i.e. more weakly
 coupled to the leads) and the electron resonances become well-isolated.
   For temperatures that are low compared  with the mean level
 spacing,  the dot's conductance  is dominated by the
 resonance that is closest to  the Fermi energy of the electrons in the leads.
 Since a tunneling event requires the addition of one electron into
the dot and a collective charging energy of $e^2/C$ (where $C$ is the
capacitance of the dot), the conductance exhibits a series of approximately
equally spaced peaks
as a function of the gate voltage \cite{reviews}.
The height of the conductance peaks, however, shows order of magnitude
fluctuations.  These fluctuations measure directly the fluctuations of
the wavefunctions in the interface region
between the dot and the leads.  Using random matrix theory (RMT),
the statistical distribution of conductance peak heights
was derived in closed form \cite{JSA92}. 
Recently, these distributions were measured in dots with single-channel
symmetric leads and several hundred electrons, for both the
case of conserved and broken time-reversal
symmetry \cite{Chang96,Marcus96}, and were found to be in good agreement
with the theoretical predictions.    The measured
 parametric correlator of  the conductance
 peak as a function of an applied magnetic field was also found to be in
agreement with the predicted correlator \cite{AA96}.  

One aspect of the data from ref. \cite{Marcus96} has remained
unexplained.   Strong correlations were observed between the heights
of adjacent peaks, in contrast to the RMT prediction of vanishing
correlations in the low temperature limit.  In this experiment
the temperature was estimated to be $0.3-0.5 \Delta$, so the
correlations could be due to deviations from the $T \ll \Delta$ results;
but then the rather good agreement of the conductance distributions
with the $T \ll \Delta$ result is somewhat puzzling.  We address these
questions here by deriving conductance peak distributions
 and peak-to-peak correlations for temperatures that are not much smaller
 than the mean-level spacing $\Delta$.  We find that, 
due to effects of the charging energy,
the corrections to the distributions are smaller than expected from
the non-interacting Landauer-B\"uttiker conductance formula \cite{LBform}.
Nonetheless significant peak-to-peak correlations are induced, 
and we will discuss their relation to experiment below.
The deviation of the finite temperature distributions from those
predicted by naive application of the Landauer-B\"uttiker formula
is somewhat surprising since on resonance the mean interaction
energy difference between the $N$ and $N-1$ particle ground states
vanishes \cite{Be91,AKL91} and indeed in the limit $T \ll \Delta$ one
obtains exactly the same distribution as in the non-interacting case
\cite{BS94,AL95}.
 Our result is derived by combining the theory of sequential
resonant tunneling in the Coulomb blockade regime \cite{Be91,AKL91} 
with the statistical assumptions of RMT.

Beenakker \cite{Be91} considered the linear response of a dot in equilibrium
with a chemical potential equal to the Fermi energy in the leads.
The electrostatic energy of the dot with  $N$ electrons is given by
$U(N) =  (Ne)^2 / 2C - N e \alpha V_g$ where  $C$ is the
capacitance of the dot, $V_g$ is the  gate voltage
 and $\alpha$
 denotes the ratio between the
plunger gate to dot capacitance and the total capacitance.
The Fermi energy
of a resonant tunneling event of the $N$-th electron into the dot is
given by the condition
 $E_F =    U(N) - U(N-1) + E_N $, where $E_N$ is the
 single-particle energy of the $N$-th level.
 Alternatively we can define an effective Fermi energy
 $\tilde{E}_F = E_F + e \alpha V_g$ for which the condition of resonant
tunneling is
 $\protect\tilde{E}_F = E_N+ \left( N - 1/2 \right) e^2/C$.
If the gate voltage is tuned to satisfy
$e \alpha V_g=  \left( N- 1/2 \right) e^2/C$ one reaches the degeneracy
condition for which  the charging energy
$ U(N) - U(N-1)$ vanishes (to order $\Delta$).  Assuming this condition,
and that $T$ is always greater than the resonance width, Beenakker showed
that the resonant conductance $G (T, \tilde{E}_F)$ can  be written as a 
weighted sum over the $T=0$ resonances $\lambda$ in the dot

\begin{eqnarray}\label{peak}
\begin{array}{ll}
 G (T,\tilde{E}_F) = \frac{e^2}{h}\, \frac{\pi \bar{\Gamma}}{4 kT} g  \;\;\; &
{\rm where} \;\;
 g = \sum_\lambda w_\lambda(T,\tilde{E}_F) g_\lambda \;.
\end{array}
\end{eqnarray}
Here  $  g_\lambda =  2  {\bar{\Gamma}}^{-1}  [ \Gamma_\lambda^l
\Gamma_\lambda^r / (\Gamma_\lambda^l + \Gamma_\lambda^r) ] $
are the level conductances,
where  $\Gamma_\lambda^{l(r)}$  is the width
of a resonance level $\lambda$ to decay into  the
 left (right) lead and $\bar{\Gamma}= \overline{\Gamma^l} +\overline{\Gamma^r}$
is the total average width.
The quantity $g_\lambda$ is dimensionless and
 temperature-independent.  $w_\lambda= w_\lambda(T, \tilde{E}_F)$ is the weight 
 with which a given resonance $\lambda$ contributes to the conductance.   
 The contribution to $w_\lambda$ from any fixed number of electrons on the dot
 is the product of the  probability that the level $E_{\lambda}$ is filled with
the dot  having that number of electrons, and the probability that there is
an empty state in the leads at the corresponding total energy \cite{Be91}:

\begin{eqnarray}\label{wBe}
w_\lambda = \sum_N 4 P_N \langle n_\lambda \rangle_{_N} \left[
1 - f\left(E_\lambda + (N- 1/2){e^2 \over C}  - \tilde{E}_F \right) \right] \;.
\end{eqnarray}
 $P_N$ is the probability that the dot has $N$ electrons,
  $\langle n_\lambda \rangle_{_N} $ is the {\it canonical} occupation of a level
 $\lambda$, and $f(\epsilon) = [1 + \exp (\epsilon/kT)]^{-1}$.
 In many experiments $T, \Delta \ll e^2/C$ and only one term in (\ref{wBe})
contributes to a given conductance peak, corresponding to
 $N_0$ electrons in the dot.  Eq. (\ref{wBe}) reduces to \cite{Be91}
 \begin{eqnarray} \label{wBe1}
w_\lambda = 4   f(\Delta F_{N_0} - \tilde{E}_F)
\langle n_\lambda \rangle_{_{N_0}} \left[
1 - f\left(E_\lambda  - \tilde{E}_F \right) \right]
\;,
\end{eqnarray}
where  $\Delta F_N = F_N - F_{N-1}$
and $F_N$ is the canonical free energy of $N$ non-interacting particles.
Here and in the following  $e \alpha V_g$ is measured relative to
$(N_0 -1/2)e^2/C$.

In the limit $T \ll \Delta$ only the central level $\lambda = N_0$ 
(denoted by $\lambda=0$ in the following) contributes
to a given conductance peak in (\ref{peak}).  Its weight $w_0$
 becomes the appropriate weight for non-interacting electrons,
which one would get by appropriate approximation of the
Landauer-B\"uttiker formula for narrow isolated resonances 

   \begin{eqnarray}\label{wLB}
w_0^{LB} = 4 kT f^\prime(E_0-\tilde{E}_F) =
 \cosh^{-2}  \left({ E_0 - \tilde{E}_F \over 2kT } \right) \;.
\end{eqnarray}

In the absence of interactions this result generalizes trivially to
the regime  where $T$ is not much smaller than $\Delta$. In this case 
 several resonances $\lambda$ contribute to (\ref{peak}) with
 weights $w_\lambda^{LB}$ obtained by replacing $E_0$ in  (\ref{wLB})
with $E_{\lambda}$.  Since the charging energy ``vanishes'' on resonance
one might have expected Eq.  (\ref{wBe}) to reduce to this form.
In fact this  only happens when $e^2/C \ll \Delta$; not in the
experimentally relevant limit $e^2/C \gg \Delta$.
If $e^2/C \rightarrow 0$, then all terms (with various number
 of electrons $N$) contribute to (\ref{wBe}). The factor $1-f$
becomes independent of $N$ and by definition
 $\sum_N P_N \langle n_\lambda \rangle_{_N} = f(E_\lambda - \tilde{E}_F)$ is 
just the grand-canonical occupation number, so that
 $w_\lambda$ reduces to  $w_\lambda^{LB}$ for all $\lambda$'s. In this
case the various manifolds of many-electron levels
 with $N_0$, $N_0 \pm 1$, $N_0 \pm 2; \ldots$   electrons on the
dot differ from each other only by an energy of order $\Delta$,
and consequently
 many of the $P_N$ are non-negligible and contribute  to $w_\lambda$. However,
 when the charging energy is large compared with $\Delta$ only two
 manifolds ($N_0$ and $N_0 -1$) are degenerate while all others are
pushed away amounts of order $e^2/C$.   Consequently,  the weights
$w_\lambda$ differ significantly from their non-interacting values
when $T \sim \Delta$.

\begin{figure}

\vspace*{5 mm}

\centerline{\epsffile{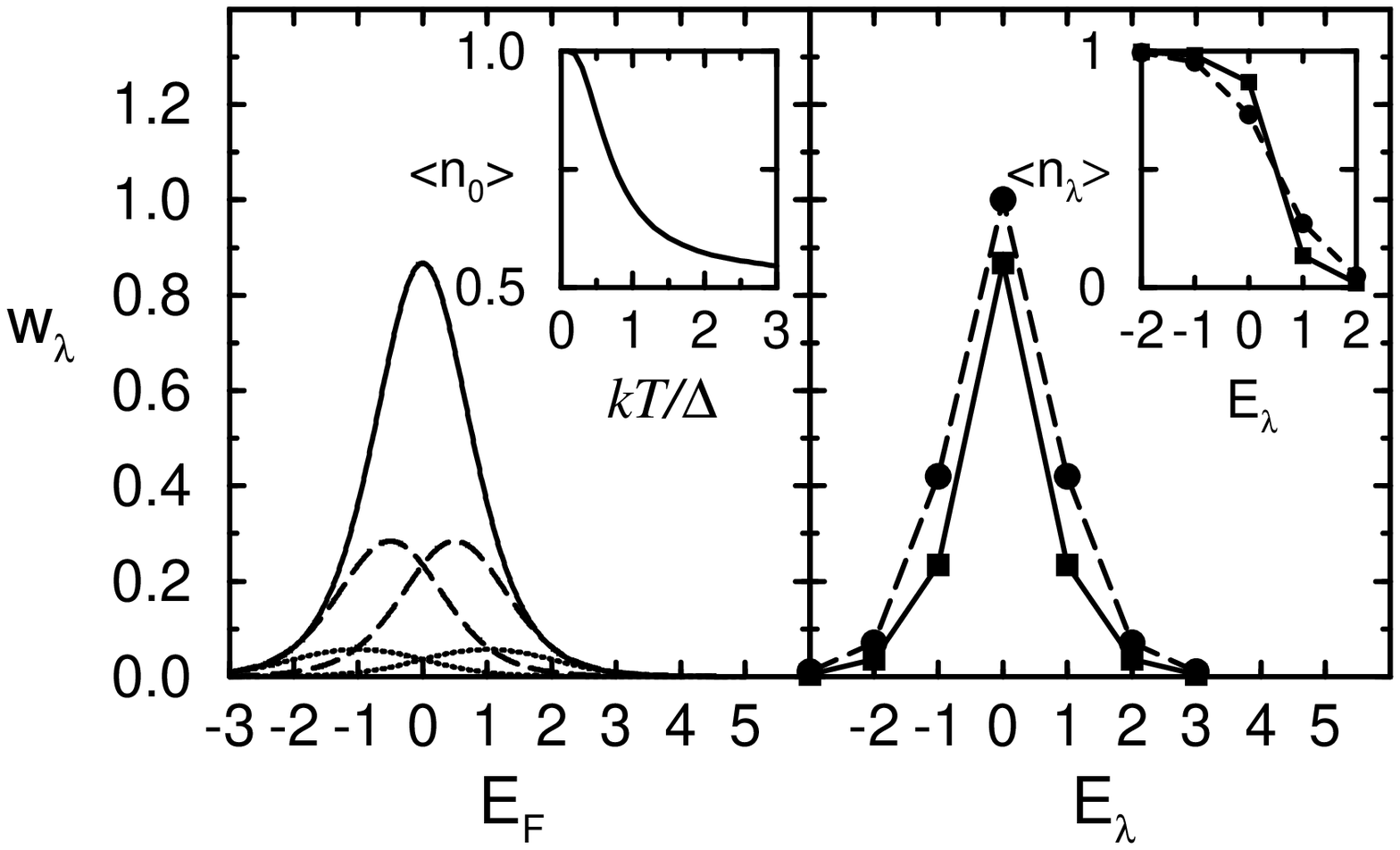}}

%\vspace{2 cm}

\caption
{ Left: The weights $w_\lambda(T, \tilde{E}_F)$ versus $\tilde{E}_F$ at
$T= 0.5 \Delta$ and for resonance levels $\lambda = 0$ (solid line),
$\pm1$ (dashed lines), $\pm2$ (dotted lines)  assuming a picket-fence
spectrum.
Right: The weights $w_\lambda$  (Eq.  (\protect\ref{wBe1}), solid squares)
 at $\tilde{E}_F = E_0$ versus $E_\lambda$ in comparison
with  the non-interacting  weights $w_\lambda^{LB}$
(solid circles).   The right inset shows the canonical occupations
$\langle n_\lambda \rangle$ (solid squares) versus $E_\lambda$ in
comparison with a Fermi-Dirac distribution (dashed line).
The left inset shows $w_0 = \langle n_0 \rangle$ as a function
 of temperature.
}

\label{fig1}

\newpage

\vspace*{5 mm}

\centerline{\epsffile{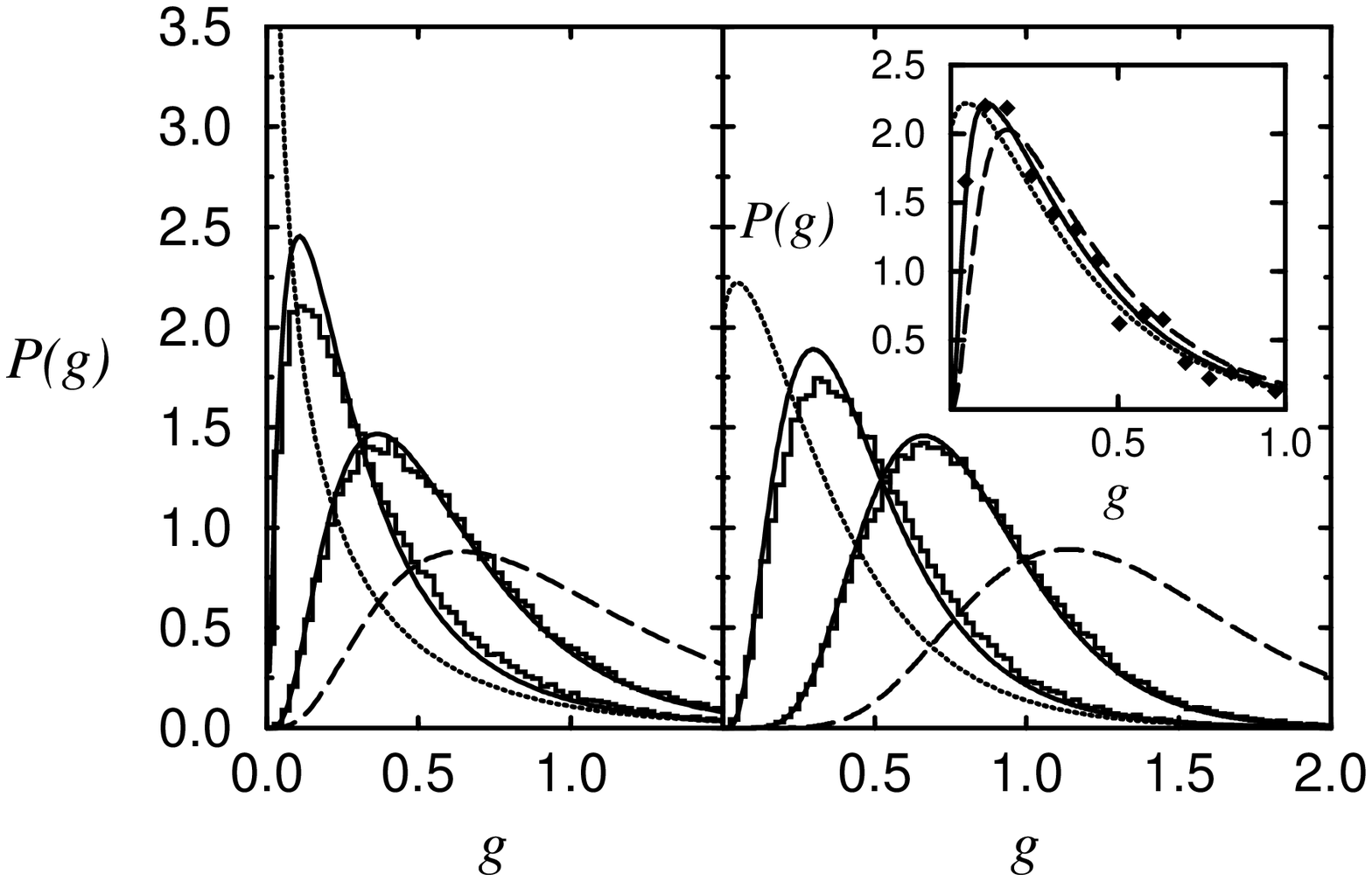}}

%\vspace{2 cm}

\caption
{ The conductance peak height distributions $P(g)$ 
  for the GOE  (left) and GUE
statistics (right).  The solid lines are the analytic
distributions (see (\protect\ref{dist})) at $T/\Delta= 0.5, 1$  for
a picket-fence spectrum using the
weights (\protect\ref{wBe1}), while the histograms describe the respective
distributions obtained  from full RMT simulations (see text). For comparison
we also show the $T \ll \Delta$ distributions (dotted lines) and the
non-interacting distributions for $T = \Delta$ (dashed lines). 
 The inset compares the
 experimental distribution in the presence
 of magnetic field \protect\cite{Marcus96} (diamonds) with the
 $T = 0.3 \Delta$ GUE distribution (solid line),
 the $T = 0.3 \Delta$ non-interacting distribution (dashed) and the
 $T \ll \Delta$ distribution (dotted).
}
\label{fig2}

\newpage

\vspace*{5 mm}

\centerline{\epsffile{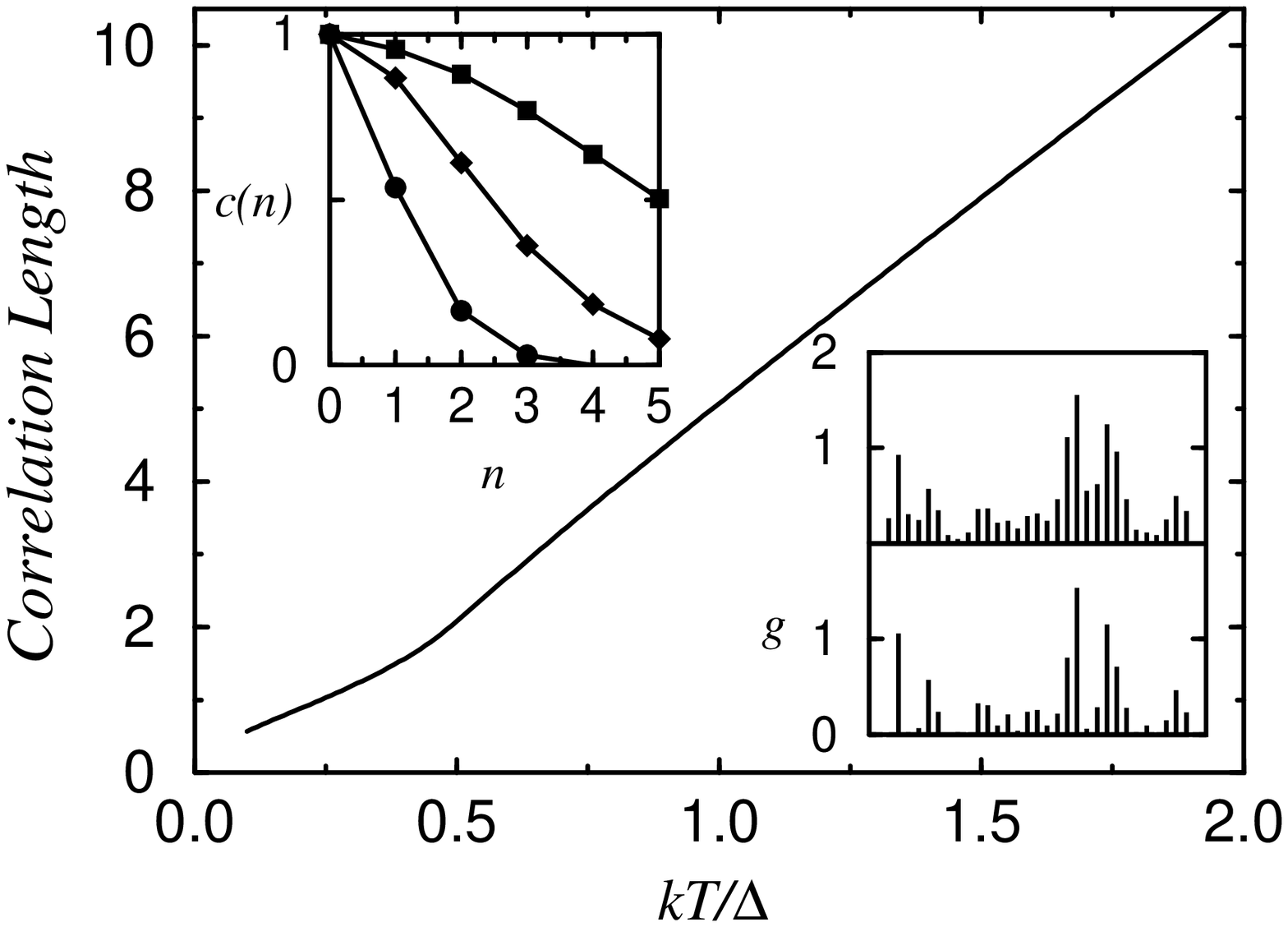}}

%\vspace{2 cm}

\caption
{ The correlation length
 (full width at half maximum) of the peak-to-peak correlator $c(n)$ as a
 function of temperature. Left inset:  $c(n)$ versus peak separation $n$ at
several temperatures $T/\Delta = 0.5, 1, 2$.  Right inset:
 a  realization of an RMT peak sequence  at $T\ll \Delta$
 (bottom part) and $T = 0.5 \Delta$ (top part).
}
\label{fig3}
\end{figure}

We now evaluate these differences quantitatively.  This
requires the calculation of the canonical quantities $F_N$ and
$\langle n_\lambda \rangle_{_N}$ in  (\ref{wBe1}), through a projection on
 a fixed number of particles $N$. This is done  in terms of  an exact
quadrature formula  \cite{quad} 
that  expresses the canonical partition function
$Z_N = e^{-F_N/T}$ in terms of grand-canonical partition functions
\begin{eqnarray}\label{partition}
 Z_N = { e^{-\beta E_g } \over N_{sp}} \sum\limits_{m=1}^{N_{sp}}
\prod_{i =1}^{N_{sp}}\left(1 + e^{-\beta |E_i - \mu|} e^{i \sigma_i \phi_m} \right)
\;.
\end{eqnarray}
Here the quadrature points are $\phi_m = 2\pi m/N_{sp}$ ($N_{sp}$ is the 
number of single-particle states), and  $\mu$ is a chemical potential chosen 
anywhere in the range $E_{N} \leq \mu < E_{N+1}$.
 $E_i - \mu$ are just the particle ($i> \mu$)
 or hole ($ i \leq \mu$) energies, and $\sigma_i=1$ for a hole and $-1$ for a 
particle.
Since the factors in each term in (\ref{partition}) decay exponentially
 as we move away from $\mu$, only a finite number of single-particle states
 $N_{sp}$ around $\mu$ are needed for an exact calculation.
The canonical occupations are calculated by a similar projection method
and differ significantly from the corresponding Fermi-Dirac
occupations  at  temperatures of order  $\Delta$ or less.
 They lie on a curve similar in shape to the a Fermi-Dirac distribution
 with a chemical potential  of
 $\mu = (E_N +E_{N+1})/2$, but with a level-dependent effective temperature
which in the vicinity of $\mu$ is smaller than the actual
temperature by almost a factor of 2 \cite{KG97}.   The inset to Fig. \ref{fig1} shows
 both distributions for $T = 0.5 \Delta$ assuming a uniformly spaced
(picket-fence) spectrum.

While the numerical calculations below include fluctuations of the 
single-particle
energies, their effect turns out to be quite small, so that
the simple picket-fence spectrum can be used to illustrate and understand the 
results.
For a picket-fence spectrum $\Delta F_N= E_N$ and $P_N = f(E_N - \tilde{E}_F)$.
In Fig.  \ref{fig1}  we show the weights $w_\lambda(T, \tilde{E}_F)$ versus
$\tilde{E}_F$ for
$T=0.5 \Delta$ for  several levels around the central level $\lambda = 0$.
The functions $w_\lambda$  become shallower and broader as we move
away from the level $\lambda = 0$, in contrast to the shape of the
Landauer-B\"uttiker weights (\ref{wLB}),  which is independent of $\lambda$.
Assuming the
conductance peaks at $\tilde{E}_F = E_0$  we also show in
Fig.  \ref{fig1}  the weights
$w_\lambda  \equiv w_\lambda(T,E_0)$ that contribute to the conductance peak
 height  vs. $E_\lambda$ \cite{symmetric}.

For the picket-fence spectrum $P_{N_0}=1/2$ and 
$w_0= \langle n_0\rangle_{_{N_0}}$ (denoted in the following by
 $\langle n_0\rangle$); whereas for all levels $\lambda
\neq 0$, the relation $w_\lambda \approx w_\lambda^{LB}/2$ holds to within
$20\%$ or better. Since $ \langle n_0 \rangle < 1 = w_0^{LB}$
the actual weights (and hence the conductance) are always smaller than
predicted by the non-interacting theory.  Hence in a rather subtle
manner the charging energy manifests itself in a suppression of the
finite temperature conductance and its fluctuations.
In the limit $T \gg \Delta$ the weight for
the central level $w_0 = \langle n_0 \rangle \to 1/2 = w_0^{LB}/2$
 (see left inset of Fig. \ref{fig1}), and
we recover the classical result \cite{Be91}
$G \approx G^{LB} /2$.  In fact we find that this limit is practically reached
at $T \approx 2 \Delta$ where $w_0$ is within $20\%$ of $1/2$.  However a
second interesting effect occurs
for $ 0.1 \Delta < T < 2 \Delta$.  In this interval
$\langle n_0 \rangle > 1/2$ and the   ratio
$w_\lambda/w_\lambda^{LB}$ is
enhanced for the central level relative to  adjacent levels.
 Thus effectively the distribution
of $g$ is less sensitive to temperature  than would be expected from
the non-interacting theory.  

To test this notion quantitatively we calculated the  conductance
distributions from Eqs.  (\ref{peak})--(\ref{wBe}) and compared them to
 the non-interacting
distributions.   A statistical theory of  the conductance in
irregularly shaped quantum dots
 was developed in Ref.  \cite{JSA92}  in the limit  $T \ll \Delta$.
 The partial  width amplitude to decay into a channel $c$ from a resonance
level  $\lambda$ is expressed
 as the projection  of the resonance wavefunction  on the
 channel wavefunction  across the interface between the dot and the lead.
   When the electron dynamics in the dot is chaotic,  the statistical 
fluctuation of the resonance wavefunction are well described by  RMT, and  the
  universal distributions of the dimensionless level conductance
 $g_\lambda$ can be derived. For  $T \ll \Delta$ and single-channel leads,
 $ P(g) = \sqrt{2/\pi g} \mbox{e}^{-2 g}$ for conserved time-reversal
 symmetry (GOE) and   $P(g) = 4 g[K_0(2g) + K_1(2g)] \mbox{e}^{-2g}$
 for broken time-reversal symmetry (GUE) \cite{JSA92,PEI93}
 (for the case of multi-channel  leads see Refs. \cite{AL95,MPA95}). 
 For temperatures comparable
 to $\Delta$, several resonances contribute to the same conductance
 peak according to (\ref{peak}).  The conductance peak is now affected
 not only by the fluctuations of the eigenfunctions but also by the statistics
 of the energy levels $E_\lambda$.  However  we
can ignore the fluctuations of the energy levels to a good approximation 
(see below) and take a picket-fence spectrum.
The conductance peak distribution can then be evaluated in closed form.
 Assuming the peak is
 positioned at  $\tilde{E}_F = E_0$, the weights $w_\lambda$ are fixed numbers,
 and only the  $g_\lambda$ fluctuate. Since in RMT different eigenfunctions
are uncorrelated we find that the characteristic function of the conductance 
peak distribution  $P(t) \equiv \langle e^{i g t} \rangle$ is  given by
\begin{eqnarray}\label{dist}
  P(t) =
\left\{  \begin{array}{ll} \prod_{\lambda}{\left( 1 - {it w_\lambda \over 2 }
 \right)^{-1/2} } & \mbox{(GOE)}  \\
\prod_\lambda {1 \over 2 (1 - {itw_\lambda \over 4})  }
\left[ 1 + { \arcsin({itw_\lambda \over 4})^{1/2} \over
 ( {itw_\lambda \over 4})^{1/2} ( 1 - {it w_\lambda \over 4})^{1/2}}
\right]  & \mbox{(GUE)}
\end{array}  \right.
\;.
\end{eqnarray}
  Fig.  \ref{fig2} shows the analytic distributions derived from (\ref{dist}) 
for both conserved (left) and broken (right) time-reversal symmetry, and for
temperatures $T= 0.1 \Delta$ (dotted lines), $0.5 \Delta$ and  $\Delta$
(solid lines).   At the lowest
temperature ($0.1 \Delta$) the distributions essentially coincide with those
 found earlier for $T \ll \Delta$, but already at $T=0.5 \Delta$ we see a
 deviation.
 The non-interacting distributions for $T = \Delta$  are shown for comparison
by the dashed lines in Fig. \ref{fig2};  they differ substantially from the 
actual
$T=\Delta$ distributions. Both the mean and the variance of the conductance
 are suppressed at a given temperature due to the presence of the charging 
energy. In the inset to Fig. \ref{fig2} we compare the measured distribution
 (solid diamonds)
 for non-zero magnetic field \cite{Marcus96} with the actual distribution we
calculate for $T = 0.3 \Delta$ (solid line), the non-interacting distribution
 at $T= 0.3 \Delta$ (dashed line) and the $T \ll \Delta$ distribution (dotted 
line).
 The finite temperature distribution describes correctly the dip observed
 in the lowest $g$ data point.  The reduced sensitivity to temperature
 (as compared with the non-interacting case) is already observed
at this low temperature.

To test our analytic approximation (\ref{dist}),
 we have done full random matrix simulations which
 include the fluctuations in the single-particle energy levels as well
as the possible fluctuations in the  peak's position. The results are shown by
 the histograms in Fig. \ref{fig2}. The largest deviations are observed for
$T = 0.5 \Delta$, but even here the analytic approximation appears to work well.

Finally we calculated the  peak-to-peak correlator (in a  peak sequence
 vs.\  gate voltage) at   finite temperature
$ c(n) =  \left(\overline{G_{N_0+n} G_{N_0}} - \overline{G_{N_0}}^2 \right)
/ \left(\overline{G_{N_0}^2} - \overline{G_{N_0}}^2 \right)$,
where $G_{N_0}$ is the conductance peak due to $N_0$ electrons on the dot.
 In the approximation that the position of each peak is fixed at
 $\tilde{E}_F = E_{N_0} +  (N_0-1/2)e^2/C$ we can express $c(n)$ in terms
of the weights $w_\lambda(N_0) \equiv w_\lambda(T, E_{N_0})$ given by
(\ref{wBe1}) for $N_0$ electrons on the dot. Since  in both the GOE and GUE
the eigenvector distribution is independent from the eigenvalues distribution,
  and $\overline{g_\lambda g_\mu} = \overline{g_\lambda^2}
\delta_{\lambda \mu} + \overline{g_\lambda}^2 (1 - \delta_{\lambda \mu} ) $
where $\overline{g_\lambda}$ and $\overline{g_\lambda^2}$ are independent of
 $\lambda$, we find

\begin{eqnarray}\label{peakcorr1}
c(n) \approx  {\overline{ \sum_\lambda w_\lambda(N_0+n) w_\lambda(N_0)} \over
\overline{\sum_\lambda w_\lambda(N_0)^2} }\;.
\end{eqnarray}
The remaining average in (\ref{peakcorr1}) is over the energy levels
 $E_\lambda$, although to a
good approximation one can again take a picket-fence spectrum.  For the latter
 case we can use the relation $w_\lambda(N_0+n) = w_{\lambda - n} (N_0)$ to 
express $c(n) \approx   \sum_\lambda w_{\lambda - n} w_\lambda /
\sum_\lambda w_\lambda^2$ in terms of the weights
$w_\lambda \equiv w_\lambda(N_0)$ of a fixed number of electrons $N_0$ on the
dot.  The top  inset of
Fig. \ref{fig3} shows $c(n)$ versus $n$ for several temperatures.
Fig. \ref{fig3} itself  shows the correlation length (defined as the  full
width at half
maximum) as a function of temperature. The increase of  the
correlations between neighboring peaks with $T/\Delta$  is in qualitative
 agreement with the experimental results which show that the peak
 distributions measured
at higher temperature \cite{Marcus96} are more strongly correlated than
those measured at lower temperatures \cite{Chang96}. The bottom inset
of  Fig. \ref{fig3} shows a sequence of conductance peaks for a particular
realization in RMT at $T \ll \Delta$ and $T = 0.5 \Delta$.
Nonetheless, the  peak series in Ref.
\cite{Marcus96} exhibits even stronger peak correlations than
we would expect for the estimated temperature of 
$T \approx  0.3 - 0.5 \Delta$.  
The origin of this enhancement of the correlations at low
temperatures is not fully understood \cite{corr}.

 In conclusion,  using RMT we have derived the finite temperature
 conductance peak  height distributions and peak-to-peak correlations
 in Coulomb blockade quantum dots. The charging energy was shown to
 affect the distribution of the conductance peaks at finite temperature
 when compared with the results of a non-interacting theory. 
 Further measurements of the mesoscopic fluctuations  of the conductance 
at  temperatures comparable to the mean level spacing  
  are necessary  for a detailed comparison with our theoretical 
results.

This work was supported in part by the Department of Energy grant
No. DE-FG-0291-ER-40608 and by NSF grant DMR-9215065.
We thank  A. Chang and C.M. Marcus for helpful conversations.

\begin{references}
\bibitem{symmetric}  Note that $w_\lambda$ are
symmetric around $\lambda=0$, which is a result of the interesting relation
$\langle n_{- m} \rangle (1-f(E_{- m}-E_0))=
\langle n_{m} \rangle (1-f(E_m - E_0))$, valid for
any spectrum which is symmetric around the central level.
\end{references}
\end{document}